\documentclass[twocolumn]{aastex631}

\usepackage{layouts}
\usepackage{amsmath}
\usepackage[nice]{nicefrac}

\usepackage{pgffor}
\usepackage{calc}

\newcommand{\comment}[1]{}
\newcommand{\diff}{{\rm d}}

\newcommand{\fevol}{f_{\rm evol}}

\def\P(#1){\Phelper#1|\relax\Pchoice(#1)}
\def\Phelper#1|#2\relax{\ifx\relax#2\relax\def\Pchoice{\Pone}\else\def\Pchoice{\Ptwo}\fi}
\def\Pone(#1){\Pr\left( #1 \right)}
\def\Ptwo(#1|#2){\Pr\left( #1 \,\middle|\, #2 \right)}
\def\Pr{\mathbf{Pr}}

\def\D{\mathcal{D}}

\def\intensityUnit{nW\,m$^{-2}$\,sr$^{-1}$}
\def\hubbleUnit{km\,s$^{-1}$\,Mpc$^{-1}$}

\newcommand{\nuInu}[1][]{
    \nu I_\nu^{\rm #1}
}

\shorttitle{The Cosmological Optical Convergence}
\shortauthors{Gréaux, Biteau \& Nievas Rosillo}

\graphicspath{{./}, {./Figures/}}

\submitjournal{ApJL}

\begin{document}

\title{The Cosmological Optical Convergence: Extragalactic Background Light from TeV Gamma Rays}

\author[0000-0003-0627-8436]{Lucas Gréaux}
\affiliation{Université Paris-Saclay, CNRS/IN2P3, IJCLab, 91405 Orsay, France}

\author[0000-0002-4202-8939]{Jonathan Biteau}
\affiliation{Université Paris-Saclay, CNRS/IN2P3, IJCLab, 91405 Orsay, France}
\affiliation{Institut Universitaire de France (IUF), France}

\author[0000-0002-8321-9168]{Mireia Nievas Rosillo}
\affiliation{Instituto de Astrofísica de Canarias, E-38205 La Laguna, Tenerife, Spain}
\affiliation{Universidad de La Laguna, Dept. Astrofísica, E-38206 La Laguna, Tenerife, Spain}

\begin{abstract}

The intensity of the extragalactic background (EBL), the accumulated optical and infrared emissions since the first stars, is the subject of a decades-long tension in the optical band. These photons form a target field that attenuates the $\gamma$-ray flux from extragalactic sources.
This paper reports the first $\gamma$-ray measurement of the EBL spectrum at $z=0$ that is purely parametric and independent of EBL evolution with redshift,  over a wavelength range from $0.18$ to $120\,\mu$m.
Our method extracts the EBL absorption imprint on more than 260 archival TeV spectra from the STeVECat catalog, by marginalizing nuisance parameters describing the intrinsic emission and instrumental uncertainties.
We report an intensity at 600\,nm of $6.9 \pm 1.9$\,\intensityUnit$\,\times\, h_{70}$, which is indistinguishable from the intensity derived from integrated galaxy light (IGL) and compatible with direct measurements taken beyond Pluto's orbit.
We exclude with $95\,\%$ confidence diffuse contributions to the EBL with an intensity relative to the IGL, $f_\mathrm{diff}$, greater than $20\,\%$ and provide a measurement of the expansion rate of the universe at $z=0$, $H_0 = 67^{+7}_{-6}$\,km\,s$^{-1}$\,Mpc$^{-1}\,\times\, (1+f_\mathrm{diff})$, which is EBL-model independent.
IGL, direct and $\gamma$-ray measurements agree on the EBL intensity in the optical band, finally reaching a cosmological optical convergence.

\end{abstract}


\section{Introduction} \label{sec:intro}

The cumulative emission from all radiating sources since the birth of the first stars forms the extragalactic background light (EBL), which is second in intensity only to the cosmic microwave background \citep[e.g.][for a review]{Driver_2021}. The broadband spectrum of the EBL is dominated by the cosmic optical background (COB, 0.1~--~8\,$\mu$m) and the cosmic infrared background (CIB, 8~--~1000\,$\mu$m). The COB and CIB consist mainly of light either directly escaping from galaxies or absorbed by dust grains and thermally radiated.
As a record of all photon production pathways since recombination, the EBL is a powerful cosmological probe and can help constrain physics beyond the Standard Model \citep[see][]{Cooray_2016, Biteau_Meyer_2022}.

Current best estimates of the EBL intensity come from the combined emission from stars, dust, and the active nuclei of galaxies, called the integrated galactic light (IGL).
Based on measured light from resolved galaxies, IGL estimates currently reach precision of 5~--~20\,\% \citep[see][]{Driver_2016, Koushan_2021}, and provide only lower bounds on the EBL, omitting contribution from low-surface-brightness and sub-threshold source populations.
In contrast, direct measurements of the EBL, which determine the cumulative light emission from both diffuse and resolved sources, provide a comprehensive view of the background intensity at the cost of contamination by foreground emissions \citep{Mattila_2019}.
Measurements from within the Solar System must, for example, account for the Zodiacal-light foreground \citep[diffuse reflection of sunlight on interplanetary dust, see][]{Hauser_2001}, which outshines the EBL by more than an order of magnitude at one astronomical unit.

Any difference between direct and IGL measurements \citep[e.g.][]{Bernstein_2007, Kawara_2017, Matsuura_2017, Lauer_2021, Lauer_2022, Symons_2023}, i.e.\ the so-called optical controversy in the visible band \citep{Driver_2021}, could be explained by 
unobserved faint galaxies \citep{Conselice_2016}, by
unresolved emissions from or around known galaxies \citep{Cooray_2012, Zemcov_2014, Matsumoto_2019},
or by physics beyond the Standard Model \citep{Bernal_2022}.
However, observations from the Hubble Space Telescope seem to disfavor at least some of these explanations \citep{Kramer_2022, Nakayama_2022}, and an excess with respect to the IGL may instead be associated with misunderstood foreground emissions.
Recent analysis by \cite{Postman_2024} of an extensive data set from the LORRI instrument aboard the New~Horizons probe \citep{Cheng_2008, Weaver_2020}, beyond Pluto's orbit where Zodiacal light is negligible, thus seems to suggest a misestimation of foreground emissions in earlier studies from the same team \citep{Lauer_2021, Lauer_2022}.

In addition to direct and IGL measurements, the COB and CIB can be measured indirectly within observational $\gamma$-ray cosmology, through their interactions with $\gamma$-rays.
Predicted by \citet{osti_4836265, Gould_1967}, observational constraints on the transparency of the universe to $\gamma$-rays were first placed by \citet{Stecker_1992}.
Two decades later, the first measurements were made using the absorption patterns induced in the $\gamma$-ray spectra of extragalactic sources at high energies \citep[HE, $0.1$--$300$~GeV,][]{FERMI_2012} and at very-high energies \citep[VHE, $0.1$--$30$~TeV,][]{HESS_2013}.
However, the most recent EBL measurements using $\gamma$-ray data only \citep{HESS_2017, MAGIC_2019, VERITAS_2019, LHAASO_2023} lack the precision to resolve the controversy \citep[see][and references therein]{Dominguez_2024}.

Following \citet{Biteau_2015} and \citet{Desai_2019}, we propose a new analysis method using a fully Bayesian framework, which we apply to the most comprehensive catalog of archival VHE observations to date, STeVECat \citep{STeVECat}, and to contemporaneous HE observations from \textit{Fermi}-LAT. This framework allows us to overcome the usual limitations of $\gamma$-ray analyses of EBL, which are related to the unknown spectra emitted by the sources \citep{Biasuzzi_2019} and to systematic uncertainties in the energy scale \citep[see e.g.][]{HESS_2013} of imaging atmospheric Cherenkov telescopes (IACTs). Our analysis also overcomes the limitations of the studies of \citet{Biteau_2015} and \citet{Desai_2019}, whose results are partially dependent on EBL evolution models (via a fixed redshift-evolution parameter and the modeled meta-observable, respectively). 

We adopt as a baseline a concordance $\Lambda$CDM model with $H_0 = 70$~\hubbleUnit, $\Omega_\mathrm{m} = 0.3$ and $\Omega_\Lambda = 0.7$. Uncertainties reported in this work correspond to 68\,\% credible intervals.

\section{Gamma-ray datasets} \label{sec:data}

\subsection{VHE $\gamma$-ray data}

We study spectra from jetted active galactic nuclei and long gamma-ray bursts published in peer-reviewed journals by ground-based $\gamma$-ray instruments from 1992 to 2021 and collected in STeVECat \citep{STeVECat}, the largest database of archival VHE spectra from extragalactic sources to date.\footnote{\url{https://zenodo.org/records/8152245}}
In STeVECat, each source is assigned a spectroscopic redshift measurement with a reliability flag from the literature review by \citet{Goldoni_2021}.
Most spectra come from the current generation of IACTs, H.E.S.S. \citep{HESS}, MAGIC \citep{MAGIC} and VERITAS \citep{VERITAS}, which observe few-degree patches of the extragalactic $\gamma$-ray sky down to ${\sim}\,50\,$GeV and up to ${\sim}\,20\,$TeV.

From STeVECat, we select non-redundant spectra with at least four flux points (excluding upper limits), from sources with a reliable redshift measurement at $z > 0.01$.
At such distance or below, substantial EBL absorption of $\gamma$-rays is only expected beyond $20$\,TeV \citep[e.g.][]{Saldana-Lopez_2021}{}{}. The threshold of four flux points per spectrum allows us to extract most of the EBL information from the $\gamma$-ray data without bias.
Our selection amounts to 268 spectra from 45 extragalactic sources (see Appendix~\ref{appendix:data_table}), going up to redshift $z=0.939$ (PKS\,1441+25).
To date, this is the most extensive VHE spectral corpus used for an EBL study: in their reference study, \citet{Biteau_2015} collected 86 spectra from 32 sources, up to redshift $z=0.287$.
Almost all sources in our data samples are jetted active galactic nuclei (mostly blazars), with the exception of three long $\gamma$-ray bursts.

\subsection{HE $\gamma$-ray data} \label{subsec:HE_data}

The Large Area Telescope (LAT) onboard the {\it Fermi Gamma Ray Space Telescope} is a pair-conversion telescope continuously observing the sky in the $\gamma$-ray band from ${\sim}\,100\,$MeV to more than $100\,$GeV \citep{atwood2009}.
As EBL absorption is negligible up to 100\,GeV for sources with redshift $z < 1$ \citep[e.g.][]{Saldana-Lopez_2021}, {\it Fermi}-LAT provides a measurement of the non-attenuated part of the spectra of $\gamma$-ray sources listed in STeVECat.
We analyzed {\it Fermi}-LAT events at energies between $100\,$MeV and $100\,$GeV with \texttt{Pass 8 SOURCE} class, within a region of interest of $15$\,deg radius around each of the STeVECat sources.
We considered a time integration window corresponding to the start and end date of the VHE observations, extended by including the 3\,h periods preceding and following the observation to ensure at least two HE snapshots of the selected region per contemporaneous STeVECat spectrum. This analysis results in 64 contemporaneous spectra.

The data were analyzed with the \textit{Fermi} Science Tools \citep{fermitools} through the high-level wrapper \texttt{enrico} \citep{sanchez2015}.
We used the latest available version of the instrument response functions ({\tt P8R3\_SOURCE\_V3}), setting the recommended zenith cuts of $90\,$deg to avoid Earth's limb contamination, and {\tt DATA\_QUAL==1 \&\& LAT\_CONFIG==1} to preserve only good quality data. We model the isotropic and Galactic diffuse background using the models tabulated in \texttt{iso\_P8R3\_SOURCE\_V3\_v1.txt} and \texttt{gll\_iem\_v07.fits}, respectively.
For each analysis, we simultaneously model all sources from the \textit{Fermi}-LAT $12$-Year Point Source Catalog \citep[4FGL-DR3,][]{abdollahi2022} within $20\,$deg of the target VHE source.
The position and extension of all sources in the sky were kept frozen to the catalog values, and we only left the normalization free for very bright sources (significance above $40\,\sigma$) in the region.
For the target source, we adopt a log-parabola spectral model, setting hard limits between $-2$ and $2$ on the spectral curvature term, to be consistent with the treatment described in Section ~\ref{sec:framework}.

\begin{figure}
    \centering
    \includegraphics[width=\linewidth]{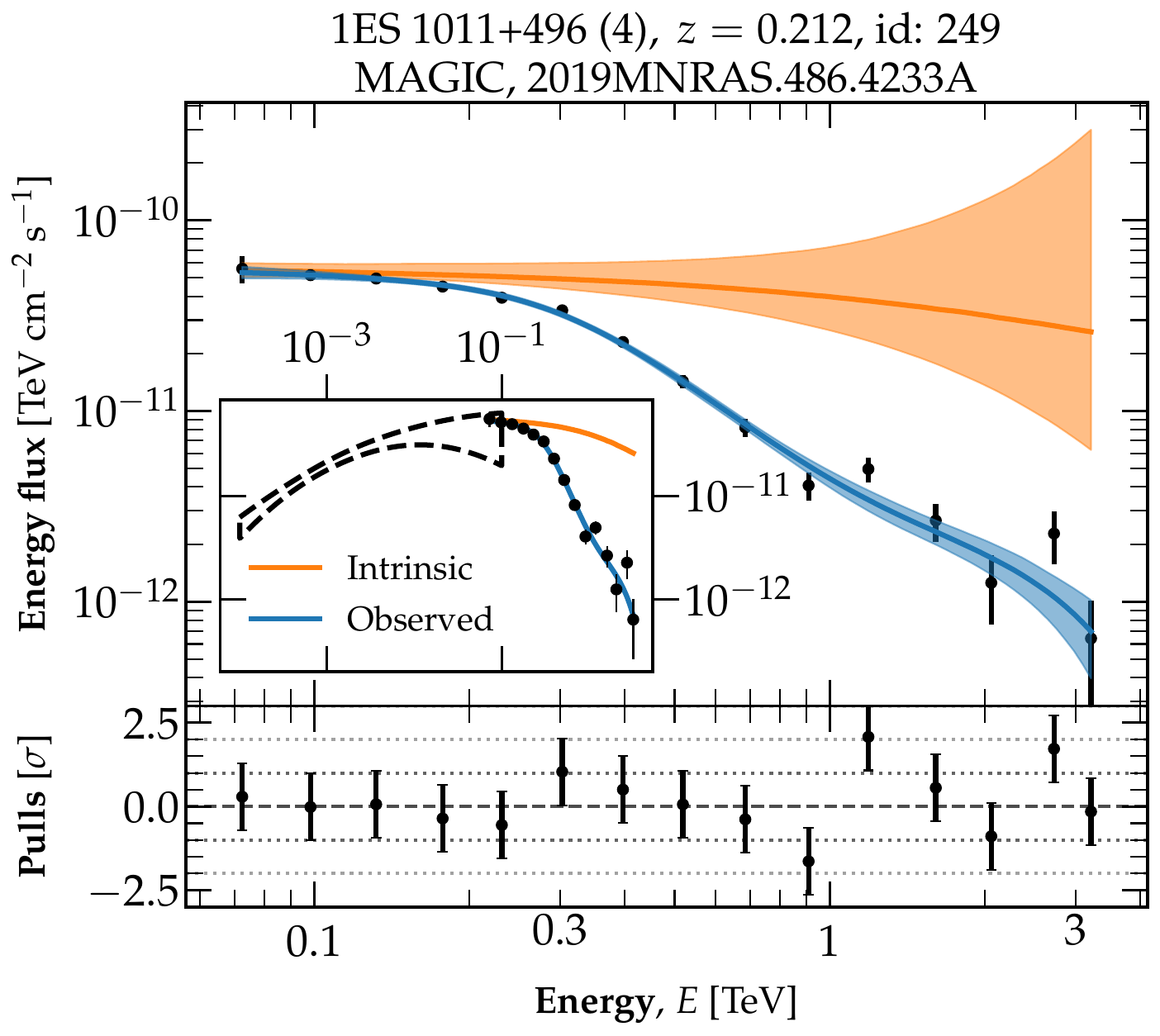}
    \caption{
    \textit{Top}: Spectrum of the source 1ES\,1011+496 observed between MJD\,56694 and MJD\,56723.
    The black points correspond to the MAGIC observation (SAO/NASA ADS reference 2019MNRAS.486.4233A), and the dashed black confidence band in the inset corresponds to the contemporaneous \textit{Fermi}-LAT observation.
    The blue curve displays the reconstructed observed spectrum, and the orange curve the reconstructed intrinsic spectrum.
    The main panel covers the VHE range, while the inset covers both HE and VHE ranges.
    Both datasets are compatible within $0.6\sigma$ at the geometric mean of the maximal HE and minimal VHE energy.
    \textit{Bottom}: Pulls, i.e. residuals normalized to the uncertainties, for the best reconstructed spectrum.
    }
    \label{fig:reconstructed_spectrum}
\end{figure}

We show in Figure~\ref{fig:reconstructed_spectrum} an exemplary spectrum used in this study, corresponding to an observation of the source 1ES\,1011+496 by the MAGIC telescopes and a contemporaneous \textit{Fermi}-LAT spectrum, shown as a dashed confidence band in the inset.
All spectra used in this work can be found in \ref{appendix:spectra}.
We use the spectral indices and curvatures reconstructed at HE as well their statistical uncertainties to establish the Bayesian priors on the intrinsic shape of the source spectra, as described in the next section.
While we assume no differential spectral shape evolution between HE and VHE samples, we do not assume that the HE flux level corresponds to the VHE flux, as blazars can exhibit large-amplitude variability on time scales as short as a few minutes \citep{2007ApJ...664L..71A}.
An overall good match is found \textit{a posteriori} between HE and VHE flux normalizations.

\section{Analysis framework} \label{sec:framework}

\subsection{Spectral model} \label{subsec:spectral_model}

The spectra emitted by VHE sources, before EBL absorption, are unknown and must be modeled.
They are best reproduced by power laws with or without intrinsic curvature and energy cut-off \citep[e.g.][]{Biteau_2015}.
The presence or absence of these spectral features can affect the reconstruction of the EBL in a frequentist framework \citep{Biasuzzi_2019}.
We overcome this difficulty by marginalizing over the intrinsic spectral parameters in a Bayesian framework. 

We model the intrinsic spectra with the general model including both curvature and energy cut-off, a log-parabola with exponential cutoff (ELP):
\begin{equation} \label{eq:ELP}
    \Phi_{\rm ELP}(E) = \Phi_0 \times e^{\eta - \Gamma \log{ \left( E/E_0 \right) } - \beta \log^2{ \left( E/E_0 \right) } - \lambda E} {\rm ,}
\end{equation}
where $E_0$ and $\Phi_0$ are fixed parameters defining a reference energy and flux.
For a VHE spectrum with energy bounds $E_{\rm min}$ and $E_{\rm max}$, we take $E_0 = \sqrt{E_{\rm min}E_{\rm max}}$, and set $\Phi_0$ as the geometric mean of the unattenuated flux at $E_{\rm min}$ and $E_{\rm max}$, using the reference attenuation from \citet{Saldana-Lopez_2021}.
The logarithmic normalization $\eta$, index $\Gamma$, curvature $\beta$ and inverse cut-off $\lambda$ are left free to vary for each spectrum.

We account for the potential bias in the energy reconstruction of VHE observatories using an energy-scale parameter $\varepsilon = \log(E/\tilde{E})$,
where $E$ is the observed energy of an event with true energy $\tilde{E}$. The joint analysis of HE and VHE spectra from the Crab Nebula yields values $-0.1 \lesssim \varepsilon \lesssim 0.1$ for different IACTs \citep{Nigro_2019}.
We model a differential spectrum observed at Earth, emitted at redshift $z$, by
\begin{equation} \label{eq:spectral_model}
    \Phi\left(E, z\right) = e^{-\tau\left(\tilde{E}, z\right)} \times \Phi_{\rm ELP}(\tilde{E}) {\rm .}
\end{equation}

The intrinsic spectral model of an observation $k$ has 5 free parameters, which we write $\theta_k \equiv \{ \eta_k, \Gamma_k, \beta_k, \lambda_k, \varepsilon_k\}$.
We write $\Theta \equiv \left\{\theta_k\right\}_k$.

\subsection{EBL absorption} \label{subsec:ebl_absorption}

The EBL-induced attenuation of the $\gamma$-ray flux observed at energy $E$ and emitted at redshift $z$ is characterized by the optical depth, integrated over redshift $z' \leq z$, comoving EBL photon energy $\epsilon'$, and comoving angle between momenta $\mu' = 1 - \cos \theta'$,
\begin{eqnarray} \label{eq:tau}
\tau{(E, z)} & = & \int^{z}_0 dz' \frac{\partial L}{\partial z'}(z')
\int^{\infty}_0 d\epsilon' \frac{4\pi}{c} \frac{\nuInu(\epsilon' , z')}{\epsilon'^2} \nonumber \\
   & & \int_{-1}^1 d\mu' \frac{1- \mu'}{2} \sigma_{\gamma \gamma}( E(1+z'), \epsilon',\mu') {\rm ,}
\end{eqnarray}
where $\nuInu$ is the EBL specific intensity and $\sigma_{\gamma \gamma}$ is the Breit–Wheeler cross-section \citep[see e.g.][]{Biteau_2015}.
For a flat $\Lambda$CDM cosmology, the distance element is
$\partial L/\partial z = c/H_0\times(1+z)^{-1}\times(\Omega_\Lambda + \Omega_\mathrm{m}(1+z)^3)^{-1/2}$.

We parametrize the EBL redshift evolution as $\nuInu(\epsilon', z') = \nuInu(\epsilon, 0) \times (1+z')^{4-\fevol}$, where $\epsilon=\epsilon'/(1+z')$ is the EBL photon energy at $z=0$. Values of $\fevol$ ranging from $1.2$ to $1.7$ have been showed by \citet{Raue_2008} and \citet{Biteau_2015} to be compatible with the EBL evolution as modeled by \citet{Kneiske_2002,Franceschini_2008, Gilmore_2012}.
In this work, we marginalize over the nuisance parameter $\fevol$ ranging from 1 and 2, to ensure the independence of the results from the choice of a specific EBL model.

Following \citet{Biteau_2015}, we parametrize the EBL spectrum at $z=0$ as a sum of Gaussian functions of $l = \ln(\lambda/\lambda_{\rm ref})$, with fixed means $\left(l_i\right)_i$ and deviation $\sigma$, leaving the amplitudes $\left(a_i\right)_i$ free to vary:
\begin{equation} \label{eq:gauss}
    \nuInu(l) = \sum_i a_i \exp\left(-\frac{(l-l_i)^2}{2\sigma^2}\right) {\rm .}
\end{equation}
We impose that the sum of two successive Gaussians of unity amplitude is equal to one in between their means, i.e.\ $l_{i+1}-l_i = \sigma \times 2\sqrt{2 \ln{2}}$.
The optical depth can be written $\tau(E, z) = \sum_i a_i t_i(E, z)$,
where the $t_i(E,z)$ are weights independent of the parametrization $\left(a_i\right)_i$ that can be computed in advance using the analytic kernel from \citet{Biteau_2015}.

We chose the means $l_i$ and deviation $\sigma$ to fully cover the $400-900\,$nm band of the LORRI instrument aboard New~Horizons: a Gaussian is centered around $\lambda_\mathrm{ref} = 600\,$nm, with $\sigma /\ln{10} = 0.15$ dex.
We model the EBL with 8 Gaussians centered at wavelengths ranging from 300\,nm to 80\,$\mu$m based on the expected reach of STeVECat.
Adding Gaussians at shorter or longer wavelengths has no impact on our results.
This modeling of the EBL has 9 parameters, which we write $a = \left\{ \{a_i\}_i, \fevol \right\}$.

\subsection{Bayesian analysis} \label{subsec:bayesian}

We search for the EBL parameters $a$ that best describe the dataset $\D = \left\{D_k\right\}_k$ of independent observations $k$ using the model from Equation~\eqref{eq:spectral_model}.
The Gaussian likelihood $\mathcal{L} \propto \P(D_k|a, \theta_k)$ quantifies the deviation between the data $D_k$ and the model of the flux:
\begin{equation} \label{eq:log_likelihood}
     \log \mathcal{L}(a, \theta_k) = -\frac{1}{2}\sum_{i \in D_k} \left(\frac{\Phi(E_i, z; a,\theta_k) - \Phi_i}{\sigma_i}\right)^2 {\rm ,}
\end{equation}
where $\Phi_i \pm \sigma_i$ is the flux observed at energy $E_i$.

This equation has 14 free parameters.
With a median count of 7.5 flux points per spectrum, a frequentist method could face the problem of over-fitting.
In the Bayesian formulation, we reconstruct the posterior distribution for each spectrum,
\begin{equation} \label{eq:local_bayes}
    \P(a, \theta_k | D_k) = \frac{\mathcal{L}(a, \theta_k)\,\pi(a)\,\pi(\theta_k)}{\int \diff a\,\diff \theta_k\,\mathcal{L}(a, \theta_k)\,\pi(a)\,\pi(\theta_k)} {\rm ,}
\end{equation}
where $\pi(a) = \P(a)$ and $\pi(\theta_k) = \P(\theta_k)$ are the priors on the EBL and spectral parameters, respectively. 

We chose weakly informative priors to minimize the \textit{a priori} knowledge on the expected EBL and spectral shape.
In the $\log{\Phi} - \log{E}$ space, Equation~\eqref{eq:ELP} is linear in parameters $\eta$, $\Gamma$, $\beta$ and $\lambda$, and we consider uniform priors centered on their expected values ($0$, $2$, $0$ and $0$, respectively).
Negative values of $\beta$ and $\lambda$ may not be physical, but they must be allowed in order to preserve reconstruction symmetry around $0$. We have confirmed through simulations that priors on these intrinsic parameters that are non-uniform or limited to positive values can lead to substantial biases in the reconstructed EBL intensities.
When contemporaneous GeV data are available, we replace the priors on $\Gamma$ and $\beta$ with a bivariate Gaussian distribution based on the \textit{Fermi}-LAT best-fit parameters and covariance (Section~\ref{subsec:HE_data}).
We adopt a Gaussian prior on the energy-scale parameter $\varepsilon$ between $-0.3$ and $0.3$ with mean $0$ and standard deviation $0.1$ (see Section~\ref{subsec:spectral_model}),
and a uniform prior on the EBL evolution parameter $\fevol$ between $1$ and $2$ (see Section~\ref{subsec:ebl_absorption}).
For each EBL normalization $a_i$, we consider a log-uniform prior between a tenth and ten times the intensity predicted by \citet{Saldana-Lopez_2021}. The results are robust to reasonable changes in EBL priors, e.g. using a uniform instead of log-uniform prior on $a_i$.

We use the Markov Chain Monte Carlo implementation \texttt{emcee} \citep{emcee} to sample the posterior distribution $\P(a, \theta_k|D_k)$ for each observation $k$.
Using Bayes' formula, the independence of the observations, and marginalizing over $\theta_k$, the posterior distribution $\P(a|\D)$ can be written
\begin{equation} \label{eq:global_bayesian}
    \frac{\P(a|\D)}{\pi(a)} = \prod_k \int \diff \theta_k\ \frac{\P({a,\theta_k}|D_k)}{\pi(a)} {\rm .}
\end{equation}

We compute the univariate distribution $\P(a_i | \D) = \int  \diff \fevol  \prod_{j \neq i} \diff a_j\ \P(a|\D)$ as the product of the univariate distribution $\P(a_i | D_k)$ for each spectrum $k$ as per Equation~\ref{eq:global_bayesian}, from which we extract the mean value and variance of $a_i$.
Similarly, we compute the covariance between $a_i$ and $a_j$ from the bivariate distribution $\P(a_i, a_j | \D)$ given the means and variances from the univariate distributions.

\section{Results and discussion} \label{sec:results}

\subsection{Reconstructed EBL intensity} \label{subsec:ebl_intensity}

We applied the framework developed in Section~\ref{subsec:bayesian} to the data presented in Section~\ref{sec:data}, choosing the \texttt{emcee} parameters to ensure the production of ten thousand independent samples.
The reconstructed spectra are shown in Figure Set~\ref{fig:reconstructed_spectrum}.
We find a good match between the spectra and contemporaneous GeV data, and report in Appendix~\ref{appendix:gev_tev_compatibility} the compatibility between HE and VHE spectra.\footnote{
Over 64 spectra, only two present a tension at more than the $3\sigma$ level, which can be attributed to a misrepresentation of the source variability between the time-integrated GeV data and the discrete TeV observation segments.
} The reconstructed intrinsic spectra are consistent with the expectations of the underlying astrophysical models, showing no detectable upward curvature or exponential increase.

\begin{figure*}[ht!]
    \includegraphics[width=\linewidth]{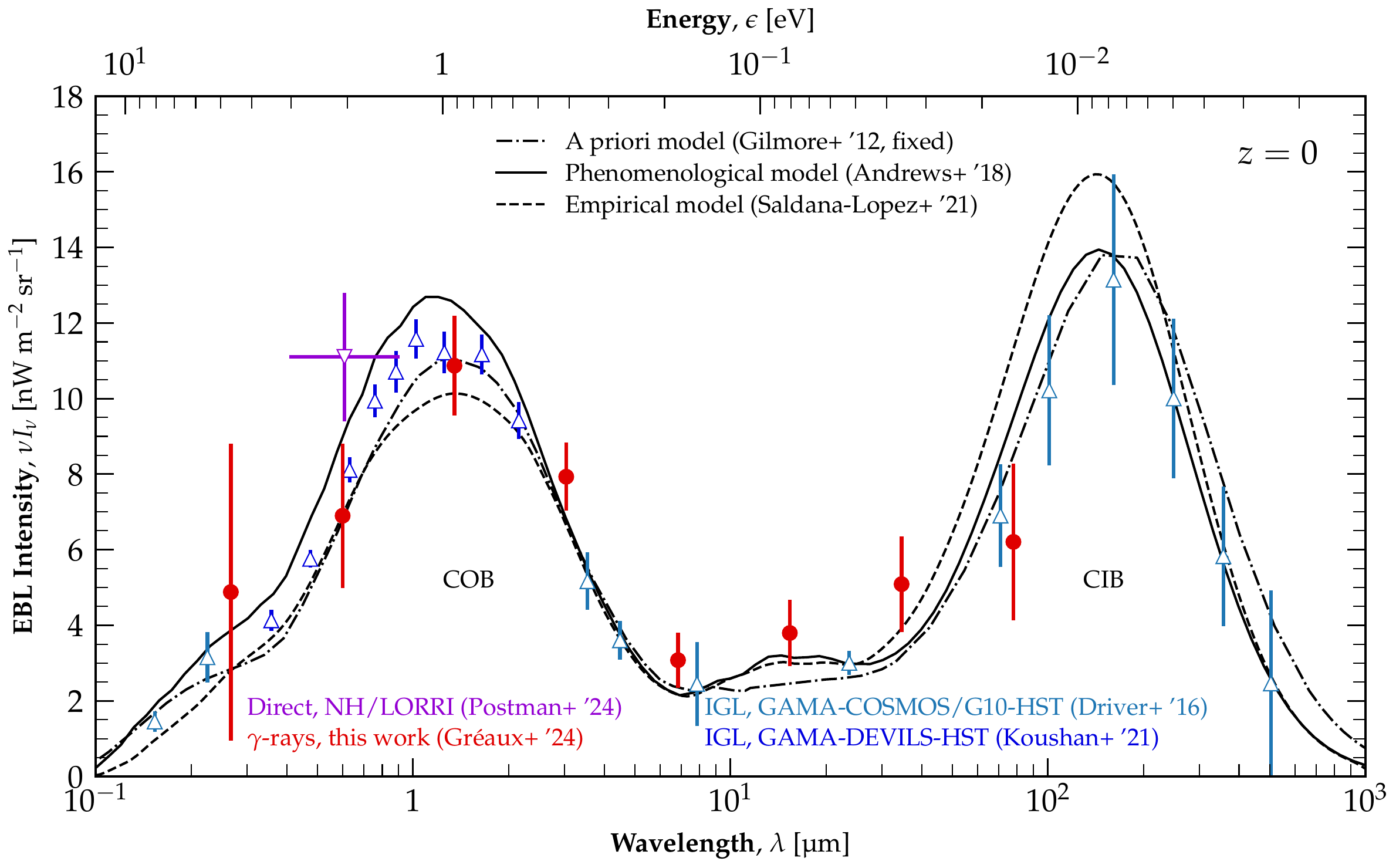}
    \caption{
    EBL intensity at $z=0$ as a function of wavelength. Indirect $\gamma$-ray measurements from this work are shown as red circles. IGL measurements from \citet{Driver_2016, Koushan_2021} are shown as upward-pointing triangles, as labeled in the figure. The direct measurement from \cite{Postman_2024} in the optical band is shown as downward-pointing triangle. The EBL models from \citet{Gilmore_2012, Andrews_2017b, Saldana-Lopez_2021} are shown as dashed-dot, solid and dashed lines, respectively. These datasets are collected in \citet{The_MM_EGAL_spectrum}.
    }
    \label{fig:ebl_flux}
\end{figure*}

The EBL specific intensity at $z=0$ resulting from the posterior distribution $\P(a|\D)$ in Equation~\eqref{eq:global_bayesian} is shown in Figure~\ref{fig:ebl_flux}, where the linear scale allows for more accurate comparisons with the IGL measurements than the logarithmic scale employed in previous studies.
The values and correlation coefficients of the reconstructed EBL intensities are given in Appendix~\ref{appendix:covariance_matrix}.
Variations on the reconstruction method of $\P(a|\D)$ from the individual samples have a negligible impact on the EBL reconstruction.
Similarly, fixing the energy-scale parameter or the EBL evolution, removing the HE prior and removing the highest-energy flux point of each spectrum have minimal effect.
Study of simulated data suggest a bias smaller than $0.4$\,\intensityUnit, which is two-to-three times smaller than the reported uncertainties. These uncertainties take full account of the statistical uncertainties and energy-scale bias of $\gamma$-ray spectral measurements, as well as the uncertainty in the intrinsic spectrum underlying each observation.

We obtain relative uncertainties on the EBL intensity of around $12\,\%$ between $1\,\mu$m and $5\,\mu$m and lower than $35\,\%$ between $0.4\,\mu$m and $120\,\mu$m.
The bluest wavelength bin is less constrained: EBL photons at $300\,$nm only induce substantial absorption of $\gamma$-rays for sources at $z > 2$, well beyond the range covered by STeVECat.
Most of the spectral corpus comes from relatively nearby sources at $z\leq0.2$, which precludes placing tight constraints on the EBL evolution with redshift. 

Our measurement is compatible with recent $\gamma$-ray attenuation measurements based either on a scaled EBL model \citep[e.g.][]{MAGIC_2019, LHAASO_2023} or on a parameterization of the EBL spectrum \citep[e.g.][]{HESS_2017, Fermi_2018, VERITAS_2019}. Our measurement is as precise as or more precise than previous VHE measurements over the entire wavelength range, even when compared to (probably less accurate) approaches that rely heavily on EBL models. Our measurement is not competitive with the HE measurement by \cite{Fermi_2018} below $400\,$nm, as expected from VHE $\gamma$-ray sources at $z<1$, but it does establish reference measurements based on $\gamma$-ray attenuation at longer wavelengths, up to $120\,\mu$m. We do not propose a joint measurement including information from the IGL as in \cite{Biteau_2015,Desai_2019}, but instead constrain the unresolved components of the EBL in the following.

\subsection{Unresolved EBL components}

The indirect $\gamma$-ray measurements shown in Figure~\ref{fig:ebl_flux} appear to be in good agreement with the IGL measurements and with the models that aim to reproduce them \citep{Gilmore_2012, Andrews_2017, Saldana-Lopez_2021}.\footnote{A comparative study of the models and their impact on astroparticle propagation will be presented in a subsequent publication.}
In particular, the IGL measurement from \citet{Koushan_2021} interpolated at 600\,nm, $\nuInu[IGL](0.6\,\mu{\rm m}) = 7.6 \pm 0.3$\,\intensityUnit, can be compared to the EBL intensity inferred from $\gamma$-ray data, $\nuInu[\gamma](0.6\,\mu{\rm m}) = 6.9 \pm 1.9 $\,\intensityUnit$\,\times\, h_{70}$, where $h_{70} = H_0 / 70$\,km\,s$^{-1}$\,Mpc$^{-1}$. 

\begin{figure}[t!]
    \includegraphics[width=\linewidth]{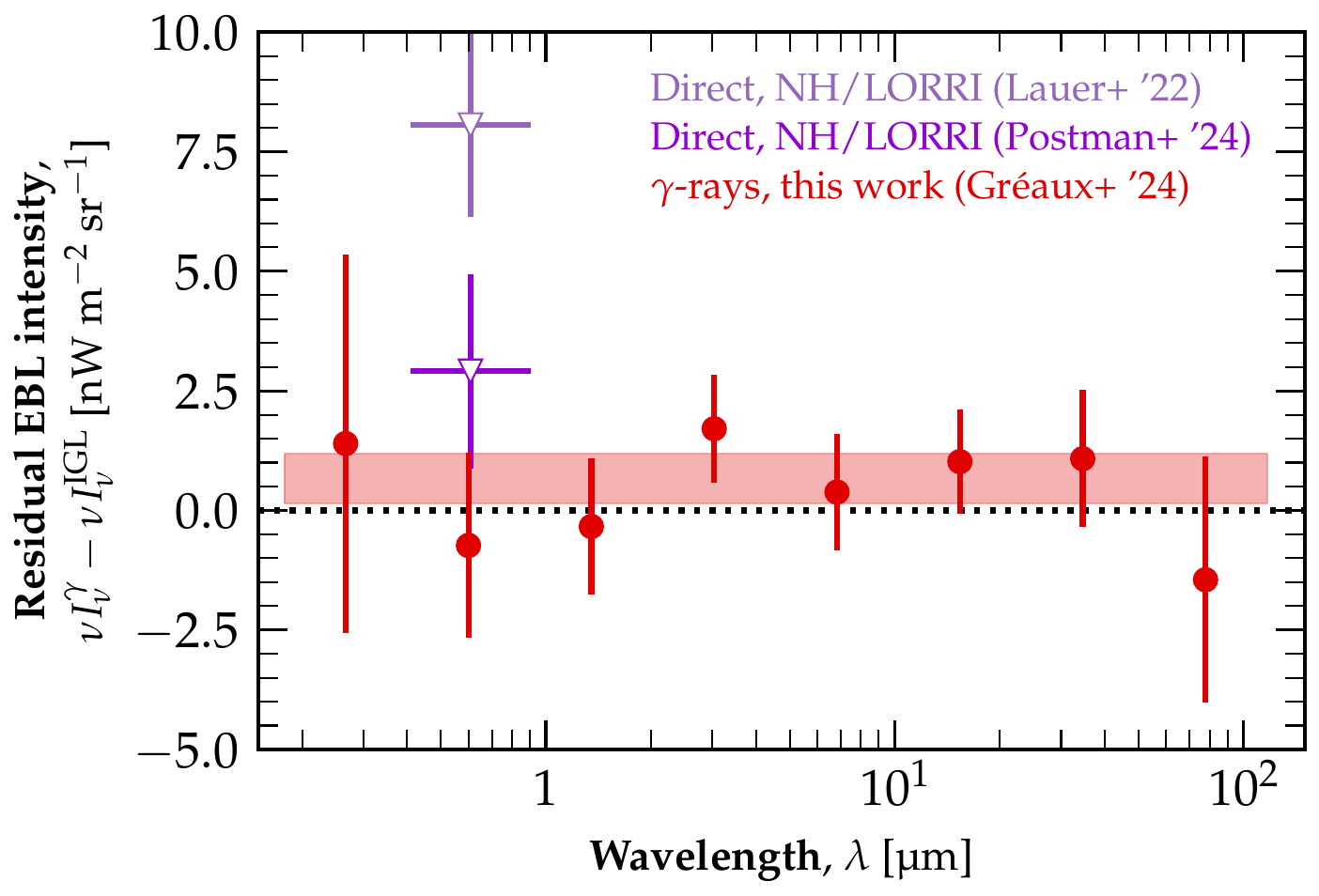}
    \caption{
    Residual EBL intensity at $z=0$ as a function of wavelength with respect to the IGL measurements from \citet{Driver_2016, Koushan_2021}.
    Results derived in this work are shown with circles for each wavelength bin.
    The horizontal band is the result of a constant fit to the residuals over the full wavelength range.
    The measurements from \citet{Lauer_2022} and \citet{Postman_2024} are shown with downward-pointing triangles, as labeled in the figure.
    }
    \label{fig:ebl_residuals}
\end{figure}

We present in Figure~\ref{fig:ebl_residuals} the residual EBL intensity with respect to IGL measurements, compared to the residual intensity from New~Horizons as determined in \cite{Lauer_2022, Postman_2024}.
Our measurement at $600$\,nm rules out an excess with respect to IGL larger than $2.5$\,\intensityUnit\ at 95\,\% confidence level.
This result is consistent with the re-analysis of data from New~Horizons by \citet{Postman_2024}, which found COB level consistent with IGL inferences.
The agreement between IGL and $\gamma$-ray data strongly suggests a foreground misestimation in studies such as \citet{Bernstein_2007, Kawara_2017, Matsuura_2017, Symons_2023}.
At wavelengths between 0.9 and 50\,$\mu$m, we can exclude residual specific intensities greater than $3.6$\,\intensityUnit\ at 95\,\% confidence level.
This value remains an order of magnitude above the expected peak intensity of the relic emission from reionization sources \citep{Cooray_2012b}.

No significant excess is found between $\gamma$-ray and IGL measurements at any wavelength.
The average excess between 200\,nm and 120\,$\mu$m, $\langle \nuInu[\gamma] - \nuInu[IGL] \rangle = 0.7\pm0.5$\,\intensityUnit, sets a limit on the fraction of non-IGL contribution to the EBL.
Assuming that diffuse, unresolved components contribute to the EBL with a spectrum similar to that of IGL and with a relative intensity $f_\mathrm{diff}$, i.e.\ $\nuInu = \nuInu[IGL] \times(1+f_\mathrm{diff})$, we exclude at 95\,\% confidence level values of $f_\mathrm{diff}$ greater than 20\,\%.
This upper limit constrains the amount of intra-halo light, whose contribution to the EBL has been estimated between 5\,\% and 30\,\% in the near infrared \citep[][]{Driver_2022, Cheng_2021}.

\subsection{Hubble constant}

The propagation of VHE $\gamma$-rays can be used to measure the expansion rate of the universe \citep{Dominguez_Prada_2013, Biteau_2015, Dominguez_2019, Dominguez_2024}.
Fixing the optical depth $\tau$ in Equation~\eqref{eq:tau} to the value inferred from $\gamma$-ray data determines the value of $\nuInu/H_0$.
The ratio of the indirect EBL measurement from $\gamma$-ray data to the IGL measurement thus reads $\nuInu[\gamma]/\nuInu[IGL] = (1+f_\mathrm{diff})/h_{70}$.
The IGL intensity is independent of the Hubble constant, as it is determined from the integral of an observed flux distribution.

Using the IGL measurements as Gaussian priors on the local intensity of the EBL, we leave the ratio $\nuInu[\gamma]/\nuInu[IGL]$ free and reconstruct
$H_0 = 67^{+7}_{-6}$\,km\,s$^{-1}$\,Mpc$^{-1}\,\times\, (1+f_\mathrm{diff})$, which is compatible with both the estimates based on cosmic microwave background observations \citep{Planck_2020} and the cosmic distance ladder \citep{Riess_2020}.
Although our uncertainties are about twice as large as those derived by \cite{Dominguez_2024}, our measurement is independent of any model of luminosity functions of galaxies contributing to the IGL at different redshifts. 

A relevant and model-independent answer to the Hubble tension that makes use of this technique will require more precise measurements of the diffuse components of the EBL and of extragalactic $\gamma$-ray spectra.

\section{Summary and conclusion} \label{sec:conclusion}

After over a decade of tension, estimates of the EBL based on galaxy surveys and direct measurements are converging in the optical band.
We present the first fully model-independent broadband spectrum of the EBL obtained from $\gamma$-ray measurements.
We used archival observations from the current generation of IACTs collected in STeVECat \citep{STeVECat} and contemporaneous \textit{Fermi}-LAT observations, which correspond to more than three times as many VHE spectra as used by previous $\gamma$-ray studies \citep{Biteau_2015, Desai_2019}.
We developed a fully Bayesian framework, which allows to circumvent the need for additional hypotheses on the emitted VHE spectra, and to marginalize over the redshift evolution of the EBL as well as over systematic uncertainties of instrumental origin.

This indirect measurement of the EBL intensity is independent of both IGL and direct measurements.
We report an intensity at 600\,nm of $6.9 \pm 1.9$\,\intensityUnit$\,\times\, h_{70}$, which is indistinguishable from the IGL measurement by \citet{Koushan_2021}, and compatible with the direct measurement from New~Horizons \citep{Postman_2024}.
IGL, direct and $\gamma$-ray measurements agree on the EBL intensity in the optical band, finally reaching a cosmological optical convergence.
The excellent agreement between the $\gamma$-ray indirect measurement and the IGL measurement over nearly three decades in wavelength leaves little room for diffuse or unresolved contributions, $f_\mathrm{diff} < 20\,\%$, and provides a measurement on the expansion rate of the universe at $H_0 = 67^{+7}_{-6}$\,km\,s$^{-1}$\,Mpc$^{-1}\,\times\, (1+f_\mathrm{diff})$.
This measurement of the Hubble constant is independent of models of evolution of the EBL and its sources.

This work is focused on the EBL and Hubble parameter at $z=0$.
A combined analysis of the present sample with $\gamma$-ray data from sources at redshifts $z>1$ observed by \textit{Fermi}-LAT could yield better constraints on other cosmological parameters such as $\Omega_\mathrm{m}$ \citep{Dominguez_2024} and on the cosmic evolution of the EBL \citep{Andrews_2017b}, which is closely linked to the cosmic star formation history \citep{Fermi_2018}.

Significant advances in this scientific field are also expected from other observatories already in operation or about to become operational.
Direct EBL measurements will benefit from a better understanding of foregrounds derived from observations by the Hubble Space Telescope \citep{Windhorst_2022} and the James Webb Space Telescope \citep{Windhorst_2023}, not to mention the constraints on reionization sources from the latter.
The galaxy counts from the large surveys by Euclid \citep[see][]{Euclid_2024} and Vera C. Rubin Observatory \citep[see][]{2019ApJ...873..111I} may bring the accuracy of IGL measurements down to the percent level.
The deployment of the Cherenkov Telescope Array Observatory \citep[CTAO, see][]{CTA_2021} will finally provide unprecedented $\gamma$-ray spectral measurements in the TeV energy range.
The stage is set for the precise measurements of the emission of all stars and galaxies since recombination.

\medskip

We thank the reviewer for constructive comments that helped to improve the quality of this manuscript.
We are also grateful to the colleagues who kindly provided comments on this manuscript, in particular Tod Lauer.
We gratefully acknowledge funding from ANR via the grant MICRO, ANR-20-CE92-0052.

\bibliography{main}{}
\bibliographystyle{aasjournal}

\appendix

\section{Gamma-ray sources} \label{appendix:data_table}

We present in Table~\ref{tab:source_table} the list of extragalactic VHE sources that have been considered in this study. We provide the commonly used name and 4FGL name of each source, its class, its redshift, its Galactic longitude and latitude, and the number of associated VHE spectra. The information from Table~\ref{tab:source_table} has been extracted from STeVECat \citep[see][]{STeVECat}. Most redshifts are spectroscopic estimates from the review of \citet{Goldoni_2021}. The STeVECat identifier of each spectrum and the corresponding bibliographic reference can be found in Figure Set~\ref{fig:reconstructed_spectrum} . 

\startlongtable
\begin{deluxetable*}{llccccc} \label{tab:source_table}
\tablecaption{Sources used in this study}
\tablewidth{700pt}
\tablehead{
    \colhead{Source name} &
    \colhead{4FGL name} &
    \colhead{Class} &
    \colhead{Redshift} &
    \multicolumn{2}{c}{Galactic coordinates [deg]} & 
    \colhead{Spectra} \\
    \colhead{} &
    \colhead{} &
    \colhead{} &
    \colhead{} &
    \colhead{Longitude} &
    \colhead{Latitude} &
    \colhead{\#}
}
\startdata
NGC\,1275                & J0319.8$+$4130 & FR-I & 0.018 & \phantom{}150.58 & \phantom{}-13.26 & \phantom{0}4 \\
IC\,310                  & J0316.8$+$4120 & AGN  & 0.019 & \phantom{}150.18 & \phantom{}-13.73 & \phantom{0}8 \\
3C\,264                  & J1144.9$+$1937 & FR-I & 0.022 & \phantom{}235.73 & \phantom{-}73.04 & \phantom{0}1 \\
Mkn\,421                 & J1104.4$+$3812 & HBL  & 0.030 & \phantom{}179.83 & \phantom{-}65.03 & \phantom{}68 \\
Mkn\,501                 & J1653.8$+$3945 & HBL  & 0.033 & \phantom{0}63.60 & \phantom{-}38.86 & \phantom{}68 \\
1ES\,2344+514            & J2347.0$+$5141 & HBL  & 0.044 & \phantom{}112.89 & \phantom{0}-9.91 & \phantom{0}5 \\
Mkn\,180                 & J1136.4$+$7009 & HBL  & 0.045 & \phantom{}131.91 & \phantom{-}45.64 & \phantom{0}1 \\
1ES\,1959+650            & J2000.0$+$6508 & HBL  & 0.047 & \phantom{0}98.00 & \phantom{-}17.67 & \phantom{}17 \\
AP\,Librae               & J1517.7$-$2422 & LBL  & 0.048 & \phantom{}340.68 & \phantom{-}27.58 & \phantom{0}1 \\
PKS\,0625$-$354          & J0627.0$-$3529 & AGN  & 0.055 & \phantom{}243.45 & \phantom{}-19.97 & \phantom{0}1 \\
I\,Zw\,187               & J1728.3$+$5013 & HBL  & 0.055 & \phantom{0}77.07 & \phantom{-}33.54 & \phantom{0}2 \\
NVSS\,J073326+515355     & J0733.4$+$5152 & EHBL & 0.065 & \phantom{}166.00 & \phantom{-}27.32 & \phantom{0}1 \\
BL\,Lacertae             & J2202.7$+$4216 & IBL  & 0.069 & \phantom{0}92.59 & \phantom{}-10.44 & \phantom{0}5 \\
PKS\,2005$-$489          & J2009.4$-$4849 & HBL  & 0.071 & \phantom{}350.37 & \phantom{}-32.60 & \phantom{0}5 \\
GRB\,190829A             & ---            & LGRB & 0.079 & \phantom{}187.68 & \phantom{}-54.99 & \phantom{0}2 \\
PMN\,J0152+0146          & J0152.6$+$0147 & HBL  & 0.080 & \phantom{}152.38 & \phantom{}-57.54 & \phantom{0}1 \\
S3\,1741+19              & J1744.0$+$1935 & HBL  & 0.084 & \phantom{0}43.84 & \phantom{-}23.34 & \phantom{0}1 \\
W\,Comae                 & J1221.5$+$2814 & IBL  & 0.102 & \phantom{}201.74 & \phantom{-}83.29 & \phantom{0}2 \\
MS\,13121$-$4221         & J1315.0$-$4236 & HBL  & 0.105 & \phantom{}307.55 & \phantom{-}20.05 & \phantom{0}1 \\
PKS\,2155$-$304          & J2158.8$-$3013 & HBL  & 0.116 & \phantom{0}17.73 & \phantom{}-52.25 & \phantom{}16 \\
B3\,2247+381             & J2250.0$+$3825 & HBL  & 0.119 & \phantom{0}98.25 & \phantom{}-18.58 & \phantom{0}1 \\
1H\,0658+595             & J0710.4$+$5908 & EHBL & 0.125 & \phantom{}157.40 & \phantom{-}25.43 & \phantom{0}1 \\
H\,1426+428              & J1428.5$+$4240 & EHBL & 0.129 & \phantom{0}77.49 & \phantom{-}64.90 & \phantom{0}3 \\
B2\,1215+30              & J1217.9$+$3007 & HBL  & 0.130 & \phantom{}188.87 & \phantom{-}82.05 & \phantom{0}9 \\
1ES\,0806+524            & J0809.8$+$5218 & HBL  & 0.138 & \phantom{}166.25 & \phantom{-}32.91 & \phantom{0}3 \\
PKS\,1440$-$389          & J1443.9$-$3908 & HBL  & 0.139 & \phantom{}325.64 & \phantom{-}18.71 & \phantom{0}1 \\
1ES\,0229+200            & J0232.8$+$2018 & EHBL & 0.139 & \phantom{}152.94 & \phantom{}-36.61 & \phantom{0}4 \\
1RXS\,J101015.9$-$311909 & J1010.2$-$3119 & HBL  & 0.143 & \phantom{}266.91 & \phantom{-}20.05 & \phantom{0}1 \\
1ES\,1440+122            & J1442.7$+$1200 & HBL  & 0.163 & \phantom{00}8.33 & \phantom{-}59.84 & \phantom{0}1 \\
H\,2356$-$309            & J2359.0$-$3038 & EHBL & 0.165 & \phantom{0}12.84 & \phantom{}-78.04 & \phantom{0}3 \\
MG4\,J200112+4352        & ---            & HBL  & 0.174 & \phantom{0}79.07 & \phantom{-0}7.11 & \phantom{0}1 \\
RX\,J0648.7+1516         & J0648.7$+$1516 & HBL  & 0.179 & \phantom{}198.99 & \phantom{-0}6.33 & \phantom{0}1 \\
PG\,1218+304             & J1221.3$+$3010 & EHBL & 0.184 & \phantom{}186.36 & \phantom{-}82.73 & \phantom{0}3 \\
1ES\,1101$-$232          & J1103.6$-$2329 & EHBL & 0.186 & \phantom{}273.19 & \phantom{-}33.08 & \phantom{0}1 \\
1ES\,0347$-$121          & J0349.4$-$1159 & EHBL & 0.188 & \phantom{}201.93 & \phantom{}-45.71 & \phantom{0}1 \\
1E\,0317.0+1835          & J0319.8$+$1845 & HBL  & 0.190 & \phantom{}165.11 & \phantom{}-31.70 & \phantom{0}1 \\
1ES\,1011+496            & J1015.0$+$4926 & HBL  & 0.212 & \phantom{}165.53 & \phantom{-}52.71 & \phantom{0}4 \\
1ES\,0414+009            & J0416.9$+$0105 & EHBL & 0.287 & \phantom{}191.81 & \phantom{}-33.16 & \phantom{0}2 \\
PKS\,1510$-$089          & J1512.8$-$0906 & FSRQ & 0.360 & \phantom{}351.29 & \phantom{-}40.14 & \phantom{0}5 \\
GRB\,190114C             & ---            & LGRB & 0.424 & \phantom{}222.47 & \phantom{}-53.08 & \phantom{0}5 \\
4C\,+21.35               & J1224.9$+$2122 & FSRQ & 0.434 & \phantom{}255.07 & \phantom{-}81.66 & \phantom{0}1 \\
3C\,279                  & J1256.1$-$0547 & FSRQ & 0.536 & \phantom{}305.10 & \phantom{-}57.06 & \phantom{0}1 \\
GRB\,180720B             & ---            & LGRB & 0.654 & \phantom{0}94.84 & \phantom{}-63.07 & \phantom{0}1 \\
B2\,1420+32              & J1422.3$+$3223 & FSRQ & 0.682 & \phantom{0}53.35 & \phantom{-}69.59 & \phantom{0}1 \\
PKS\,1441+25             & J1443.9$+$2501 & FSRQ & 0.939 & \phantom{0}34.56 & \phantom{-}64.70 & \phantom{0}3
\enddata
\tablecomments{The third column provides the object class. LGRB stands for long-duration gamma-ray burst, FR-I for radio galaxy of Fanaroff-Riley Type 1,  AGN for active galactic nucleus of unknown subclass, FSRQ for flat-spectrum radio quasar. LBL, IBL, HBL, EHBL stand for low-, intermediate-, high-, and extremely-high-synchrotron peak BL Lac.}
\end{deluxetable*}

\section{Compatibility between GeV data and TeV reconstruction}
\label{appendix:gev_tev_compatibility}

We present in Table~\ref{tab:fermi_table} the list of spectra for which a contemporaneous \textit{Fermi}-LAT observation has been considered in this study.
We provide the global and per-source identifiers of each spectrum, the commonly used name and 4FGL name of the corresponding source, the \textit{Fermi}-LAT observation period considered and corresponding livetime, and the tension between the GeV data and the TeV reconstruction at the midpoint of the HE and VHE energy bands.

\startlongtable
\begin{deluxetable}{cllccc} \label{tab:fermi_table}
\tablecaption{Spectra with contemporaneous \textit{Fermi}-LAT observations}
\tablewidth{700pt}
\tablehead{
    \colhead{Spectrum Id.} & 
    \colhead{Source name} &
    \colhead{4FGL name} &
    \colhead{Observation period} &
    \colhead{HE Livetime [h]} & 
    \colhead{$\sigma_\textrm{HE,VHE}$}
}
\startdata
    \phantom{00}1 \phantom{0}(1) & NGC\,1275       & J0319.8$+$4130 & 55043.875 - 56992.125 & $\phantom{}46758$ & $0.0$ \\
\phantom{00}2 \phantom{0}(2) & NGC\,1275       & J0319.8$+$4130 & 57631.875 - 57813.125 & $\phantom{0}4350$ & $1.4$ \\
\phantom{00}3 \phantom{0}(3) & NGC\,1275       & J0319.8$+$4130 & 57753.775 - 57754.205 & $\phantom{000}10$ & $0.2$ \\
\phantom{00}4 \phantom{0}(4) & NGC\,1275       & J0319.8$+$4130 & 57754.765 - 57756.155 & $\phantom{000}33$ & $0.3$ \\
\phantom{00}5 \phantom{0}(1) & IC\,310         & J0316.8$+$4120 & 55122.875 - 55241.125 & $\phantom{0}2838$ & $2.0$ \\
\phantom{0}51 \phantom{}(38) & Mkn\,421        & J1104.4$+$3812 & 54849.875 - 54983.125 & $\phantom{0}3198$ & $0.6$ \\
\phantom{0}55 \phantom{}(42) & Mkn\,421        & J1104.4$+$3812 & 56307.131 - 56307.408 & $\phantom{0000}7$ & $0.0$ \\
\phantom{0}57 \phantom{}(44) & Mkn\,421        & J1104.4$+$3812 & 56312.047 - 56312.379 & $\phantom{0000}8$ & $0.0$ \\
\phantom{0}58 \phantom{}(45) & Mkn\,421        & J1104.4$+$3812 & 56334.955 - 56335.285 & $\phantom{0000}8$ & $0.0$ \\
\phantom{0}74 \phantom{}(61) & Mkn\,421        & J1104.4$+$3812 & 56776.015 - 56776.455 & $\phantom{000}11$ & $0.3$ \\
\phantom{0}76 \phantom{}(63) & Mkn\,421        & J1104.4$+$3812 & 57756.894 - 57757.406 & $\phantom{000}12$ & $> 5$ \\
\phantom{0}77 \phantom{}(64) & Mkn\,421        & J1104.4$+$3812 & 57784.877 - 57785.265 & $\phantom{0000}9$ & $0.0$ \\
\phantom{0}78 \phantom{}(65) & Mkn\,421        & J1104.4$+$3812 & 57787.914 - 57788.192 & $\phantom{0000}7$ & $0.0$ \\
\phantom{0}79 \phantom{}(66) & Mkn\,421        & J1104.4$+$3812 & 57788.900 - 57789.224 & $\phantom{0000}8$ & $1.0$ \\
\phantom{}102 \phantom{}(21) & Mkn\,501        & J1653.8$+$3945 & 54682.875 - 55850.125 & $\phantom{}28014$ & $0.2$ \\
\phantom{}103 \phantom{}(22) & Mkn\,501        & J1653.8$+$3945 & 54906.875 - 55004.125 & $\phantom{0}2334$ & $0.9$ \\
\phantom{}104 \phantom{}(23) & Mkn\,501        & J1653.8$+$3945 & 54912.875 - 55038.125 & $\phantom{0}3006$ & $1.5$ \\
\phantom{}105 \phantom{}(24) & Mkn\,501        & J1653.8$+$3945 & 54937.875 - 54955.125 & $\phantom{00}414$ & $3.7$ \\
\phantom{}106 \phantom{}(25) & Mkn\,501        & J1653.8$+$3945 & 54937.875 - 54955.125 & $\phantom{00}414$ & $2.9$ \\
\phantom{}107 \phantom{}(26) & Mkn\,501        & J1653.8$+$3945 & 54937.875 - 54955.125 & $\phantom{00}414$ & $1.2$ \\
\phantom{}108 \phantom{}(27) & Mkn\,501        & J1653.8$+$3945 & 54951.875 - 54955.125 & $\phantom{000}78$ & $0.6$ \\
\phantom{}109 \phantom{}(28) & Mkn\,501        & J1653.8$+$3945 & 54972.875 - 54974.125 & $\phantom{000}30$ & $0.6$ \\
\phantom{}110 \phantom{}(29) & Mkn\,501        & J1653.8$+$3945 & 55850.875 - 55887.125 & $\phantom{00}870$ & $1.2$ \\
\phantom{}113 \phantom{}(32) & Mkn\,501        & J1653.8$+$3945 & 56031.875 - 56033.125 & $\phantom{000}30$ & $0.2$ \\
\phantom{}124 \phantom{}(43) & Mkn\,501        & J1653.8$+$3945 & 56086.875 - 56088.125 & $\phantom{000}30$ & $1.0$ \\
\phantom{}126 \phantom{}(45) & Mkn\,501        & J1653.8$+$3945 & 56093.875 - 56095.125 & $\phantom{000}30$ & $0.0$ \\
\phantom{}129 \phantom{}(48) & Mkn\,501        & J1653.8$+$3945 & 56395.054 - 56395.348 & $\phantom{0000}7$ & $0.0$ \\
\phantom{}131 \phantom{}(50) & Mkn\,501        & J1653.8$+$3945 & 56421.017 - 56421.334 & $\phantom{0000}8$ & $0.0$ \\
\phantom{}165 \phantom{}(10) & 1ES\,1959+650   & J2000.0$+$6508 & 56033.875 - 56079.125 & $\phantom{0}1086$ & $1.1$ \\
\phantom{}166 \phantom{}(11) & 1ES\,1959+650   & J2000.0$+$6508 & 56066.875 - 56068.125 & $\phantom{000}30$ & $0.0$ \\
\phantom{}167 \phantom{}(12) & 1ES\,1959+650   & J2000.0$+$6508 & 57331.875 - 57344.125 & $\phantom{00}294$ & $2.6$ \\
\phantom{}170 \phantom{}(15) & 1ES\,1959+650   & J2000.0$+$6508 & 57552.875 - 57554.125 & $\phantom{000}30$ & $0.3$ \\
\phantom{}171 \phantom{}(16) & 1ES\,1959+650   & J2000.0$+$6508 & 57569.875 - 57571.125 & $\phantom{000}30$ & $1.0$ \\
\phantom{}173 \phantom{0}(1) & AP\,Librae      & J1517.7$-$2422 & 55325.875 - 55689.125 & $\phantom{0}8718$ & $1.5$ \\
\phantom{}174 \phantom{0}(1) & PKS\,0625$-$354 & J0627.0$-$3529 & 56231.875 - 56292.125 & $\phantom{0}1446$ & $0.2$ \\
\phantom{}176 \phantom{0}(2) & I\,Zw\,187      & J1728.3$+$5013 & 57306.875 - 57328.125 & $\phantom{00}510$ & $0.1$ \\
\phantom{}179 \phantom{0}(2) & BL\,Lacertae    & J2202.7$+$4216 & 55739.375 - 55740.625 & $\phantom{000}30$ & $0.0$ \\
\phantom{}180 \phantom{0}(3) & BL\,Lacertae    & J2202.7$+$4216 & 57187.875 - 57201.125 & $\phantom{00}318$ & $0.6$ \\
\phantom{}181 \phantom{0}(4) & BL\,Lacertae    & J2202.7$+$4216 & 57187.875 - 57189.125 & $\phantom{000}30$ & $1.8$ \\
\phantom{}182 \phantom{0}(5) & BL\,Lacertae    & J2202.7$+$4216 & 57666.039 - 57666.416 & $\phantom{0000}9$ & $0.3$ \\
\phantom{}187 \phantom{0}(5) & PKS\,2005$-$489 & J2009.4$-$4849 & 54972.875 - 55014.125 & $\phantom{00}990$ & $0.2$ \\
\phantom{}191 \phantom{0}(1) & S3\,1741+19     & J1744.0$+$1935 & 55295.875 - 55707.125 & $\phantom{0}9870$ & $0.4$ \\
\phantom{}209 \phantom{}(15) & PKS\,2155$-$304 & J2158.8$-$3013 & 56402.875 - 56601.125 & $\phantom{0}4758$ & $0.2$ \\
\phantom{}210 \phantom{}(16) & PKS\,2155$-$304 & J2158.8$-$3013 & 56804.875 - 56817.125 & $\phantom{00}294$ & $1.4$ \\
\phantom{}215 \phantom{0}(3) & H\,1426+428     & J1428.5$+$4240 & 55196.875 - 56657.125 & $\phantom{}35046$ & $2.1$ \\
\phantom{}216 \phantom{0}(1) & B2\,1215+30     & J1217.9$+$3007 & 55561.875 - 55600.125 & $\phantom{00}918$ & $0.7$ \\
\phantom{}218 \phantom{0}(3) & B2\,1215+30     & J1217.9$+$3007 & 56900.875 - 57234.125 & $\phantom{0}7998$ & $1.5$ \\
\phantom{}220 \phantom{0}(5) & B2\,1215+30     & J1217.9$+$3007 & 57265.875 - 57600.125 & $\phantom{0}8022$ & $1.9$ \\
\phantom{}222 \phantom{0}(7) & B2\,1215+30     & J1217.9$+$3007 & 57631.875 - 57965.125 & $\phantom{0}7998$ & $1.5$ \\
\phantom{}226 \phantom{0}(2) & 1ES\,0806+524   & J0809.8$+$5218 & 55567.875 - 55623.125 & $\phantom{0}1326$ & $0.0$ \\
\phantom{}227 \phantom{0}(3) & 1ES\,0806+524   & J0809.8$+$5218 & 55615.875 - 55617.125 & $\phantom{000}30$ & $0.0$ \\
\phantom{}228 \phantom{0}(1) & PKS\,1440$-$389 & J1443.9$-$3908 & 55984.875 - 56074.125 & $\phantom{0}2142$ & $0.6$ \\
\phantom{}232 \phantom{0}(4) & 1ES\,0229+200   & J0232.8$+$2018 & 55926.875 - 57387.125 & $\phantom{}35046$ & $1.2$ \\
\phantom{}242 \phantom{0}(3) & PG\,1218+304    & J1221.3$+$3010 & 54800.875 - 54952.125 & $\phantom{0}3630$ & $1.1$ \\
\phantom{}245 \phantom{0}(1) & 1E\,0317.0+1835 & J0319.8$+$1845 & 54731.875 - 55485.125 & $\phantom{}18078$ & $0.8$ \\
\phantom{}248 \phantom{0}(3) & 1ES\,1011+496   & J1015.0$+$4926 & 55609.875 - 56070.125 & $\phantom{}11046$ & $1.5$ \\
\phantom{}249 \phantom{0}(4) & 1ES\,1011+496   & J1015.0$+$4926 & 56693.875 - 56723.125 & $\phantom{00}702$ & $0.6$ \\
\phantom{}252 \phantom{0}(1) & PKS\,1510$-$089 & J1512.8$-$0906 & 54909.875 - 54918.125 & $\phantom{00}198$ & $0.1$ \\
\phantom{}256 \phantom{0}(5) & PKS\,1510$-$089 & J1512.8$-$0906 & 57538.875 - 57539.125 & $\phantom{0000}6$ & $0.6$ \\
\phantom{}262 \phantom{0}(1) & 4C\,+21.35      & J1224.9$+$2122 & 55364.783 - 55365.056 & $\phantom{0000}7$ & $0.4$ \\
\phantom{}265 \phantom{0}(1) & B2\,1420+32     & J1422.3$+$3223 & 58868.175 - 58870.425 & $\phantom{000}54$ & $0.9$ \\
\phantom{}266 \phantom{0}(1) & PKS\,1441+25    & J1443.9$+$2501 & 57129.875 - 57135.625 & $\phantom{00}138$ & $1.7$ \\
\phantom{}267 \phantom{0}(2) & PKS\,1441+25    & J1443.9$+$2501 & 57132.875 - 57140.125 & $\phantom{00}174$ & $0.9$ \\
\phantom{}268 \phantom{0}(3) & PKS\,1441+25    & J1443.9$+$2501 & 57135.375 - 57139.625 & $\phantom{00}102$ &\ \ $0.8$
\enddata
\end{deluxetable}

\section{Reconstructed EBL intensities and covariance matrix}\label{appendix:covariance_matrix}

We report in Table~\ref{tab:ebl_intensities} the EBL intensities obtained by applying the Bayesian framework discussed in this work to the $\gamma$-ray data from STeVECat and the contemporaneous \textit{Fermi}-LAT observations. We report in Table~\ref{tab:ebl_covariances} the correlation matrix associated to these intensities.

\begin{deluxetable}{r|cccc}[h!]
\tablecaption{Reconstructed EBL intensities at redshift $z=0$.\label{tab:ebl_intensities}}
\tablewidth{0pt}
\tablehead{
    \colhead{} & \colhead{$\lambda$} &  \colhead{$\lambda_{\rm min}$} & \colhead{$\lambda_{\rm max}$} & \colhead{$\nuInu(\lambda)$} \\
    \colhead{} & \colhead{$\mu$m} & \colhead{$\mu$m} & \colhead{$\mu$m} & \colhead{\intensityUnit}
}
\startdata
1 & $0.27$ & $0.18$ & $0.40$ & $\phantom{0}4.9 \pm 3.9$\\
2 & $0.60$ & $0.40$ & $0.90$ & $\phantom{0}6.9 \pm 1.9$\\
3 & $1.35$ & $0.90$ & $2.02$ & $          10.9 \pm 1.3$\\
4 & $3.04$ & $2.02$ & $4.56$ & $\phantom{0}7.9 \pm 0.9$\\
5 & $6.83$ & $4.56$ & $10.3$ & $\phantom{0}3.1 \pm 0.7$\\
6 & $15.4$ & $10.3$ & $23.1$ & $\phantom{0}3.8 \pm 0.9$\\
7 & $34.6$ & $23.1$ & $51.9$ & $\phantom{0}5.1 \pm 1.3$\\
8 & $77.8$ & $51.9$ & $117 $ & $\phantom{0}6.2 \pm 2.1$\\
\enddata
\end{deluxetable}

\begin{deluxetable}{cccccccc}
\tablecaption{Correlation matrix of the different EBL intensities reconstructed in this work.\label{tab:ebl_covariances}}
\tablewidth{0pt}
\tablehead{
    \colhead{1} & \colhead{2} & \colhead{3} & \colhead{4} & \colhead{5} & \colhead{6} & \colhead{7} & \colhead{8}
}
\startdata
\phantom{-}1.00 & \phantom{-}0.11 & \phantom{-}0.19 & \phantom{-}0.01 &           -0.22 &           -0.13 &           -0.13 &           -0.18 \\
   $\ \ \cdots$ & \phantom{-}1.00 & \phantom{-}0.46 & \phantom{-}0.37 & \phantom{-}0.02 &           -0.14 &           -0.02 &           -0.01 \\
   $\ \ \cdots$ &    $\ \ \cdots$ & \phantom{-}1.00 & \phantom{-}0.51 & \phantom{-}0.47 &           -0.13 &           -0.15 &           -0.02 \\
   $\ \ \cdots$ &    $\ \ \cdots$ &    $\ \ \cdots$ & \phantom{-}1.00 & \phantom{-}0.57 & \phantom{-}0.24 &           -0.05 &           -0.02 \\
   $\ \ \cdots$ &    $\ \ \cdots$ &    $\ \ \cdots$ &    $\ \ \cdots$ & \phantom{-}1.00 & \phantom{-}0.64 & \phantom{-}0.43 & \phantom{-}0.12 \\
   $\ \ \cdots$ &    $\ \ \cdots$ &    $\ \ \cdots$ &    $\ \ \cdots$ &    $\ \ \cdots$ & \phantom{-}1.00 & \phantom{-}0.67 & \phantom{-}0.39 \\
   $\ \ \cdots$ &    $\ \ \cdots$ &    $\ \ \cdots$ &    $\ \ \cdots$ &    $\ \ \cdots$ &    $\ \ \cdots$ & \phantom{-}1.00 & \phantom{-}0.55 \\
   $\ \ \cdots$ &    $\ \ \cdots$ &    $\ \ \cdots$ &    $\ \ \cdots$ &    $\ \ \cdots$ &    $\ \ \cdots$ &    $\ \ \cdots$ & \phantom{-}1.00 \\
\enddata
\end{deluxetable}

\section{Reconstructed spectra}\label{appendix:spectra}

\begin{figure}[!h]
    \centering
    \includegraphics[width=\textwidth]{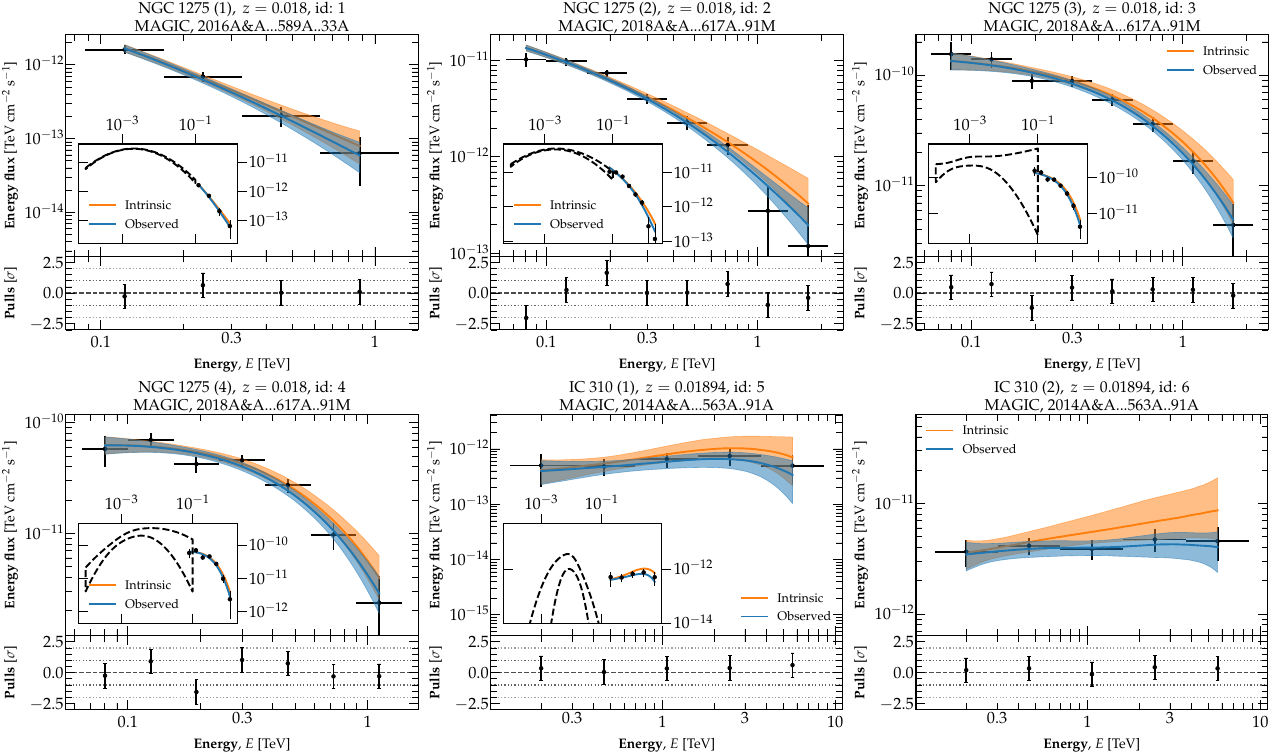}
    \caption{
    Reconstructed spectra, 1 of 23.
    For each spectrum, the main panel covers the VHE range.
    The black points correspond to the archival VHE observation obtained from STeVECat.
    The blue curve displays the reconstructed observed spectrum, and the orange curve the reconstructed intrinsic spectrum.
    When contemporaneous \textit{Fermi}-LAT observations were obtained, they are shown with a dashed black confidence band in the inset of the figures, which covers the HE range.
    The compatibility between the HE data and reconstructed VHE spectra are reported in Table~\ref{tab:fermi_table}.
    The bottom panel correspond to the pulls, i.e. residuals normalized to the uncertainties, for the best reconstructed spectrum.
    }
    \label{fig:spectra}
\end{figure}

We present in Figure~\ref{fig:spectra} and in the subsequent figures the best-fit spectral models reconstructed for each observation.

\newcounter{z}

\foreach \n in {1,...,22}{
    \edef\twodigits{\ifnum\n<10 0\n\else\n\fi}
    \setcounter{z}{\n+1}
    
    \begin{figure}[t!]
        \centering
        \includegraphics[width=\textwidth]{spectra_\twodigits.pdf}
        \caption{
            Reconstructed spectra, \arabic{z} of 23 (see Figure~\ref{fig:spectra}).
        }
    \end{figure}
}

\end{document}